%% file: main.tex
\documentclass[twocolumn, superscriptaddress, secnumarabic, floatfix, nofootinbib, nobalancelastpage,longbibliography]{revtex4-1}

\def\TitleOfPaper{Lies My Teacher Told Me About Density Functional Theory: Seeing Through Them with the Hubbard Dimer}
\def\TheHeading{\textcolor{TITLECOL}{DFT on Hubbard Dimer}}
\def\preprintlinklocation{http://dft.uci.edu + PAPER REF}

\input Burke_template
\input KieronsMacros
\usepackage{tcolorbox}
\renewcommand{\epsilon}{\varepsilon}

\begin{document}
\sf
\coloredtitle{\TitleOfPaper}

\coloredauthor{Kieron Burke}
\affiliation{Department of Physics \& Astronomy, University of California, Irvine, CA 92697}
\affiliation{Department of Chemistry, University of California, Irvine, CA 92697}

\coloredauthor{John Kozlowski}
\affiliation{Department of Chemistry, University of California, Irvine, CA 92697}

\date{\today}

\begin{abstract}
Most realistic calculations of moderately correlated materials begin with a ground-state density functional theory (DFT) calculation. While Kohn-Sham DFT is used in about 40,000 scientific papers each year, the fundamental underpinnings are not widely appreciated. In this chapter, we analyze the inherent characteristics of DFT in their simplest form, using the asymmetric Hubbard dimer as an illustrative model. We begin by working through the core tenets of DFT, explaining what the exact ground-state density functional yields and does not yield. Given the relative simplicity of the system, almost all properties of the exact exchange-correlation functional are readily visualized and plotted. Key concepts include the Kohn-Sham scheme, the behavior of the XC potential as correlations become very strong, the derivative discontinuity and the difference between KS gaps and true charge gaps, and how to extract optical excitations using time-dependent DFT. By the end of this text and accompanying exercises, the reader will improve their ability to both explain and visualize the concepts of DFT, as well as better understand where others may go wrong. This chapter appears in the book \href{https://urldefense.com/v3/__https://www.cond-mat.de/events/correl21/manuscripts/burke.pdf__;!!CzAuKJ42GuquVTTmVmPViYEvSg!OIIaXCf5McnT4tsM9QXtH_nLZOSLCCwgayrKe-7-fMn_kV-oE3covg4xEBqJLm7m2kzkObWaOU-6Aah1xoKXvZ5_$}{Autumn School on Correlated Electrons: Simulating Correlations with Computers} (2021) prepared by Forschungszentrum Jülich.
\end{abstract}

\maketitle
\tableofcontents

\sec{Introduction}

Density functional theory (DFT) is an extremely sophisticated approach to many-body problems \cite{M00,B11}. It must be among the most used and least understood of all successful theories in physics. Currently, about 50,000 papers each year report results of Kohn-Sham (KS) DFT calculations \cite{PGB15}, including room temperature superconductors under high pressure \cite{PEE20}, heterogeneous catalysis at metal surfaces and for nanoparticles \cite{NASB11}, understanding the interior of Jupiter and exoplanets \cite{Zeng19}, studying how ocean acidification affects the seabream population \cite{VRBHH19}, and even which water to use when making coffee \cite{HCC14}.

But much of modern condensed matter physics involves using model Hamiltonians to study strongly correlated systems, where understanding new phenomena is considered far more important than generating accurate materials-specific properties \cite{L11,S08}. In fact, our standard diagrammatic approach (expansions in the strength of the electron-electron coupling) is hard-wired into all our descriptions of such many-body phenomena, be it the fractional quantum Hall effect \cite{TSG82} or the Kondo effect (even when perturbation theory fails, we still think of resummed diagrams) \cite{K64}.   

Because DFT is {\em logically} subtle, without requiring much mathematical gymnastics (although they are available for those that enjoy them \cite{L83}) or skill with summing Feynman diagrams, and because DFT is entirely different from the standard approach, most of what you may have learned is hopelessly confused or simply downright untrue.  Hence the title of this article, taken from a popular book on history \cite{Loewen08}. For example, any conflation of the KS scheme with traditional mean-field theory is a dire mistake, and should be avoided at all costs.

This chapter is primarily designed to explain essential concepts of DFT to theorists more familiar with standard many-body theory and perhaps more experienced in dealing with strongly correlated systems. It should also prove useful for anyone performing DFT calculations on weakly correlated systems, who might be wondering where things go wrong as correlations grow stronger. Additionally, the Hubbard dimer is a wonderful teaching tool for basic concepts, as so many of its exact results can be derived analytically.

The first use of this material came in a conversation between KB and Duncan Haldane at a meeting sponsored by the US Department of Energy.  Duncan asked KB to explain this DFT business, and he suggested the dimer as the minimal relevant model. After 45~minutes of tough argument, Haldane said ``That's the first time I've ever really understood this Kohn-Sham scheme. Thanks.'' Within 2 years, he was awarded a share in a Nobel Prize in physics \cite{H17}. While correlation is not causation, Haldane did not win his share until {\em after} he understood KS-DFT with the aid of this simple model!

However, it is important to note that the benefits of this type of analysis are not solely limited to those working in theoretical physics. In the fields of theoretical chemistry and material science, for instance, where ground-state electronic energies are often required to be extremely accurate \cite{LS95,FP07,VSKSB19}, there has been growing technological interest in the study of both chemically complex and strongly correlated materials \cite{HWC11,OG91}. This chapter was partly designed with these fields in mind, serving as a resource for any computational scientist who wishes to better comprehend the limitations of their computational methods. Throughout this text, there will be various highlighted sections dedicated to examples, exercises, and key concepts to aid the reader in applying what is learned in this study to their own endeavors.

There are now a huge number of diverse introductions to DFT, with many different perspectives. These include a simple tutorial for anyone with knowledge of quantum mechanics \cite{BW13}, a very long online textbook with lots of nasty problems \cite{ABC}, a many-body introduction \cite{DG12}, and even video lectures \cite{YouTube}. But this chapter is specifically aimed at explaining the most essential concepts, and why strongly correlated systems are more challenging in DFT\@. All the Hubbard material appears in two long review articles, one on the ground state theory \cite{CFSB15} and a second on linear-response TDDFT \cite{CFMB18}.  The Hubbard dimer has been recently used to explore effects in other aspects of DFT, such as magnetic DFT \cite{U18}, ensemble DFT \cite{SB18}, and thermal DFT \cite{SB20}.

\begin{tcolorbox}
\textbf{Takeaway:} DFT appears deceptively simple to understand. It is much trickier than people realize. This chapter provides a unique explanation of basic ideas using a simple model.
\end{tcolorbox}

\ssec{Background}

We work in the non-relativistic non-magnetic Born-Oppenheimer approximation, using Hartree atomic units ($e^2=\hbar=m_e=1$). The Hamiltonian for the electrons is simple and known exactly
\begin{equation}
    \hat H = \hat T + \hat V\ee + \hat V \!,
\label{ham1}
\end{equation}
where $\hat T$ is their kinetic energy, $\hat V\ee$ is the electron-electron Coulomb repulsion, and $\hat V$ is the one-body potential, equal to a sum of Coulomb attractions to the ions in an isolated molecule or solid. We let $N$ be the number of electrons.

A first-principles approach to this problem is to feed a computer a list of nuclear types and positions and, following a recipe, it spits out various properties of the electronic system. In quantum chemistry \cite{SO82}, the recipe is called a model chemistry \cite{OPW95,BM07} if both the method (e.g.~Hartree-Fock) and the basis set are specified.

We contrast this with traditional approaches in condensed matter \cite{GY19}. Often a model Hamiltonian is written down, hoping that it describes the dominant physical effects. For most interesting problems, standard approaches to solving this Hamiltonian will fail, i.e., be hopelessly inadequate or require near-infinite computer resources. An inspired approximation may be found that works well enough, and so the underlying physics can be explained. Well enough will usually mean that with good estimates of the model parameters, qualitative and even semi-quantitative agreement is found with key properties of interest.

Each of these are excellent approaches, especially for the purposes they were designed for. Modern DFT calculations of weakly correlated materials (and molecules) are of the first-prin\-ci\-ples type, and often yield atomic positions within 1-2 hundredths of an \AA ngstrom and phonon frequencies within 10\%, without any materials-specific input, an impossibility with a simple model Hamiltonian. On the other hand, with standard approximations, DFT calculations always fail whenever a bond such as H$_2$ is stretched, and correlations become strong \cite{CMY08}. Even simple Mott-Hubbard physics is beyond such methods (and we shall see why in this chapter), or Kondo physics (but see Reference \cite{JSK20}).

But more and more of modern materials research requires the intelligent application of both approaches, and many methods, such as DFT+U \cite{HFGC14} or dynamical mean field theory (DMFT) \cite{APKA97,KSHO06,V11,P11} are being developed to bridge the gap. Many of the materials of greatest practical interest to energy research (such as for batteries \cite{HWC11} or photovoltaics \cite{OG91}) include a moderate level of correlation that require a pure DFT approach to be enhanced, by adding vital missing ingredients of the physics.   

The US and Britain are friends `separated by a common language' \cite{Wilde87}. This is essentially true of the mass of confusion between traditional many-body theory and DFT\@. In DFT, we use the same words as in MBT, but giving them different meanings, simply because we enjoy confusing folks.

Finally, we mention an intermediate Hamiltonian between the dazzling complexity of the real physical and chemical world and the beautiful simplicity of the Hubbard model. A great challenge to studying the effects of strong correlation has been the difficulty in producing highly accurate benchmark data. Molecular electronic structure calculations are much simpler than materials calculations, and quantum chemistry has long been able to provide highly accurate answers for many small molecules at or near equilibrium \cite{BM07}, as well as the complete binding energy curves of others \cite{H86}. But this is much harder to do for materials. Recent illustrations of this difficulty are the careful bench-marking of model Hamiltonians (such as an $8{\times}8$ Hubbard lattice) using highly accurate many-body solvers \cite{KC12}, the amount of computation needed to find an accurate cohesive energy of the benzene crystal \cite{YHUMSC14}, and the celebration of merely being able to agree on approximate DFT results with a variety of solid-state codes \cite{Sci16}.

To overcome this difficulty, about 10 years ago, a mimic of realistic electronic structure calculations was established \cite{SWWB12}. This mimic uses potentials that are defined continuously in space (i.e., not a lattice model) but are one-dimensional. In fact, ultimately, a single exponential was chosen \cite{BSWBW15}, whose details mimic those of the popular soft Coulomb potential. With about 20 grid points per `atom', standard density-matrix renormalization (DMRG) methods \cite{W92,W93} could then rapidly produce extremely accurate ground-state energies and densities for chains of up to about 100 atoms \cite{SWWB12}. By living in 1D, not only is DMRG very efficient, but the thermodynamic limit (of the number of atoms going to infinity with fixed interatomic spacing) is also reached much more quickly than in 3D\@. Moreover, the parameters were chosen so that standard density functional approximations, such as the local density approximation \cite{KS65}, succeeded and failed in ways that were qualitatively similar to those in the real world \cite{WSBW12}. We will refer to this 1D laboratory for further demonstration of some of the simple results shown in this chapter.

\begin{tcolorbox}
\textbf{Takeaway:} DFT is ideally suited to produce useful accuracy for ground-state energetics of realistic Hamiltonians. Many-body theory is more often used to produce approximate answers to model Hamiltonians, and often focuses on response properties. Both are useful in their own fields and, increasingly, interesting problems require input from both.
\end{tcolorbox}

\ssec{Hubbard dimer}
\index{Hubbard dimer}
The Hubbard model (in 1, 2 or 3D) \cite{H63} is the standard model for studying the effects of strong correlation on electrons. By default, it implies an infinite periodic array of sites. For our demonstration, we simply need two sites. We have $N=2$ and the ground-state is always a singlet. The Hamiltonian (in 2nd quantization) is
\begin{equation}
    \hatH = -t \sum_{\sigma} \left(\hat{c}_{1\sigma}^{\dagger}\hat{c}_{2\sigma} + h.c. \right) + U \sum_{i} \hat{n}_{i\uparrow\,}\hat{n}_{i\downarrow} + \sum_i v_i \hat{n}_i \,.
\label{ham2}
\end{equation}
The kinetic term is just hopping between the sites, and is the discretization of the kinetic operator on the lattice, with the diagonal elements set to 0. The electron-electron repulsion is just an onsite $U\!$, while the one-body operator is just an on-site potential, $v_1$ and $v_2$.

In this chapter, we imagine a world in which Eq.~\eqref{ham1} is replaced by Eq.~\eqref{ham2}, i.e., as if the many-body problem to be solved is simply that of Eq.~\eqref{ham2}. So, for us, the Hubbard dimer is {\em not} an approximation to anything. We will choose the values of $U\!$, $t$, and $v_i$ as we wish, to explore various regimes in the model. Any question concerning the origins of these values in terms of realistic orbitals and matrix elements is irrelevant to our work here. 

Since a constant in the potential is just a shift in the energy, we set $v_2{\,=\,}{-}v_1$ and use the parameter $\dd v{\,=\,}v_2{-}v_1$ as the sole determinant of the potential of our system. Similarly, with $N{\,=\,}2$, $n_2{\,=\,}N{-}n_1$, and we use $\dd n{\,=\,}n_2{-}n_1$ as the single parameter characterizing the ground-state density. Thus ground-state DFT in this model is simply site-occupation function theory (SOFT) and density functionals are replaced by simple functions of a single variable,~$\dd n$. Finally, we choose $t{\,=\,}1/2$ and report all variables in units of $2t$, as one can scale all energies by a constant.

\begin{figure}[t]
 \centering
 \vspace{-1ex}
 \includegraphics[width=0.49\textwidth]{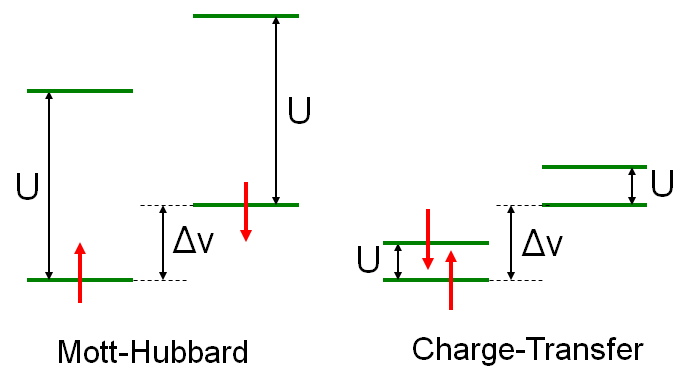}
 \vspace{-2ex}
 \caption{Two distinct regimes of the asymmetric Hubbard dimer. On the left, the charging energy is much greater than the difference in on-site potentials, and the left- and right-occupation numbers are similar. On the right, the situation is reversed, and the occupation on the left is much greater than that of the right. Reproduced from Ref. \cite{CFSB15}.}
 \label{MBdiag}
\end{figure}

Different physics appears depending on the ratio of $U$ to $\dd v$, i.e., on-site repulsion versus inhomogeneity, see Fig.~\ref{MBdiag}. When $U{\,\gg\,}\dd v$, the system is strongly correlated, with both site occupations close to $1$, despite any inhomogeneity. For $\dd v \gg U\!$, the system is weakly correlated, and the on-site $U$ is insufficient to stop one occupation becoming much greater than the other.

For those with a chemical inclination, this is a minimal basis model for a diatomic with $2$ electrons (with some matrix elements and orbital overlap ignored). For H$_2$, $\dd v=0$, but $t$ decreases as the separation between the nuclei is increased, so that $U$ (in units of $2t$) grows exponentially.
The ground-state is close to a single Slater determinant near equilibrium ($U{\ll}1$), so that Hartree-Fock (HF) is a reasonable approximation. But $U{\gg}1$ when very stretched, so that the ground-state is now a Heitler-London wavefunction, and (restricted) HF is very poor. The highly unsymmetric case corresponds to HeH$^+$, where both electrons reside on the He side, as long as $\dd v$ remains larger than $U$ as the bond is stretched.

\begin{figure}[t!]
 \centering
 \vspace{-3ex}
 \includegraphics[width=0.5\textwidth]{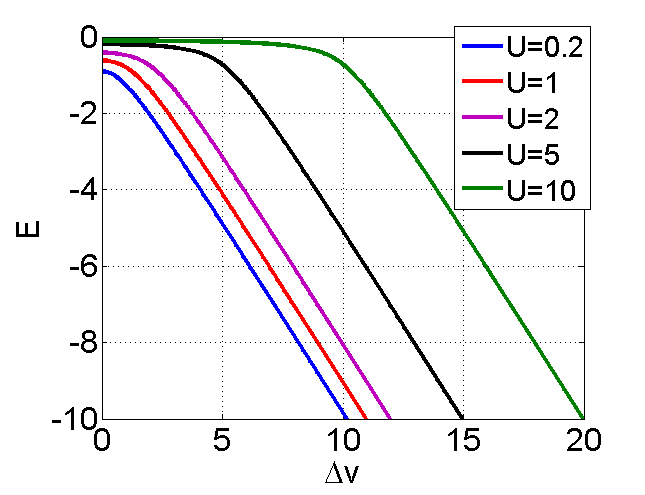}
 \vspace{-2ex}
 \caption{Exact ground-state energy of the Hubbard dimer as a function of $\dd v$ for several values of $U\!$.\, The qualitative behavior changes as $\dd v$ passes through $U\!$. Reproduced from Ref. \cite{CFSB15}.}
 \label{gsenergy}
\end{figure}

There are well-known analytic solutions for all states of the 2-site Hubbard model and the behavior of the ground-state energy \cite{CFSB15} is shown in Fig.~\ref{gsenergy}. Simple limits include the symmetric case
\begin{equation}
    \hspace{1ex} E = -\sqrt{1 + (U/2)^2} + U/2,~~~~~\dd n=0\hspace{6ex}\mbox{SYM}
\label{sym}
\end{equation}
An expansion of the square root in the symmetric case in powers of $U$ has a radius of convergence of 2, while the opposite expansion in $1/U$ has a radius of 1/2. Thus there is a well-defined critical point at $U{\,=\,}2$, below which perturbation in the electron-electron coupling strength converges, i.e., the system is weakly correlated, and above which it is strongly correlated. Another simple limit is the non-interacting (tight-binding) case ($U{\,=\,}0$)
\begin{equation}
    E = -\sqrt{1 + \dd v^2},~~~~\dd n = -2\,\frac{\dd v}{\sqrt{1 + \dd v^2}}\hspace{4ex}\mbox{($U{=}0$)} 
\label{dn_N2_U0}
\end{equation}
which is given by the blue curve in the figure. We see from the figure that, on a broad scale, $E \approx -(\dd v - U)\, \Theta (\dd v -U)$. Explicit formulas exist for all the excited-state energies, wavefunctions, and densities also. Approximations in many different limits are given in the many appendices of Reference \cite{CFSB15}.

\begin{figure}[t]
 \centering
 \includegraphics[width=0.5\textwidth]{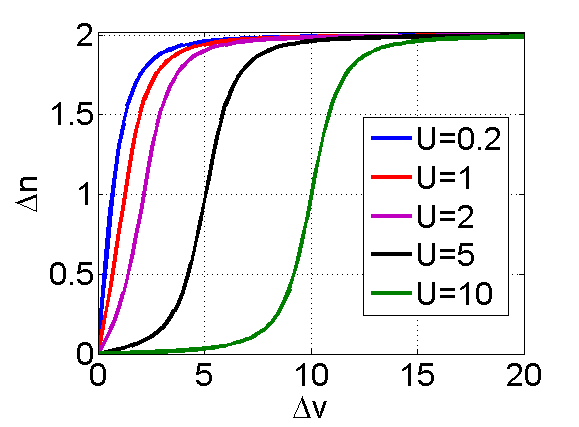}
 \caption{Ground-state occupation of the Hubbard dimer as function of $\dd v$ for several values of $U\!$. Reproduced from Ref. \cite{CFSB15}.}
 \label{dndv}
\end{figure}
We can also extract any other property we wish from the analytic solution, such as the one-electron density (here the occupations). Fig.~\ref{dndv} shows the ground-state density as a function of $\dd v$ for several values of $U\!$.\, For any $U\!$,\, $n_2=n_1$ when $\dd v=0$. The blue line is essentially the tight-binding solution. In that case, as $\dd v$ increases, the occupation difference rapidly increases towards 2. Then, as we turn on $U\!$,\, this increase becomes less and less rapid. By the time $U$ reaches 10, the occupations remain close to balanced until $\dd v$ becomes close to 10, when (on the scale of $\dd v$), it rapidly flips to close to 2.

\begin{tcolorbox}
\textbf{Takeaway:} We take the 2-site Hubbard model as our Hamiltonian, and apply DFT concepts directly to it. Here, it is not a simple model for a more realistic Hamiltonian. Analytic solutions are trivial, and we can plot any properties we wish.
\end{tcolorbox}

\sec{Density functional theory}
\index{density functional theory}
We have now defined the machinery required to understand the central theorems of DFT through the lens of the Hubbard dimer. The theorems discussed in this section, like their real-space counterparts, are exact and apply directly to ground-state calculations (we will cover time-dependent DFT later). Most DFT calculations are used to determine the ground-state electronic energy of a system, or more specifically, determine the energy of a system as a function of nuclear coordinates. In this section, we will discuss the underlying principles of these calculations by examining their role at the most fundamental level, in their simplest form.  

The Hohenberg-Kohn theorem\cite{HK64} is actually three theorems in sequence.  These were proved in a simple proof-by-contradiction argument based on the Rayleigh-Ritz variational principle for the wavefunction. Later, the more direct and more general constrained search approach was given by Levy \cite{L82} and Lieb \cite{L83}.

\begin{figure}[t!]
 \centering
 \includegraphics[width=0.47\textwidth]{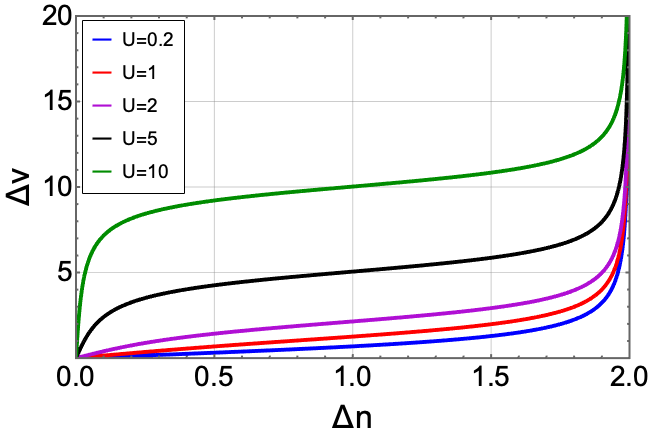}
 \caption{Ground-state potential difference as a function of $\dd n$ for several values of $U$.}
 \label{dvdn}
\end{figure}

\ssec{Hohenberg-Kohn I}
\index{Hohenberg-Kohn!Theorem I}
HKI proves that the (usual) map of $\dd v \to \dd n$ is invertible, i.e., $\dd n$ is a single-valued function of $\dd v$ for a given $U\!$.\, This is obvious from Fig.~\ref{dndv} (and its inversion, Fig.~\ref{dvdn}), and in the TB case
\begin{equation}
    \hspace{13ex} \dd v = \frac{\dd n}{\sqrt{4 - \dd n^2}} \hspace{13ex} \mbox{($U{=}0$)}.
\label{TBdv}
\end{equation}

\noindent Fig.~\ref{dvdn} is simply Fig.~\ref{dndv} drawn sideways, i.e., with x and y axes reversed. Clearly, for any given value of $U\!$,\, there is a unique $\dd v$.  

A much-stated (but often out of context) corollary of this is that {\em all} properties of the system are (implicitly) functionals of $n_1$. While this is true, almost all research in DFT focuses on the ground-state energy functional, because it is so useful, and we have few useful approximations for others (e.g., for the first excited-state energy, but see discussion in TDDFT section). Recently, machine learning methods have been trained to find some of these other functionals \cite{MCG20,MFG21}.

\ssec{Hohenberg-Kohn II}
\index{Hohenberg-Kohn!Theorem II}
HKII states that the function below exists and is independent of $\dd v$:
\begin{equation}
    F_{\sss U}(n_1) = \min_{\mathrm{\Psi} \rightarrow n_1} {\langle\mathrm{\Psi}|\hat{T} + \hat{V}\ee|\mathrm{\Psi}\rangle}
    = \max_{\dd v} \bigg\{ E(\dd v)- \dd v \dd n /2 \bigg\} .
\label{HK2}
\end{equation}
where the minimum is over all antisymmetrized normalized 2-electron wavefunctions whose occupation of site 1 is $n_1$. The middle expression is the constrained search definition due to Levy \cite{L79}. The rightmost form is due to Lieb \cite{L83}. Either definition works here. This $F_{\sss U}$ functional was termed universal by HK, by which they simply meant that it does not depend on the $\dd v$ of your given system, i.e., it is a pure density functional. The phrase, often appearing in the literature, that $F$ is a universal functional, is not meaningful.

\begin{figure}[t]
 \centering
 \vspace{-2ex}
 \includegraphics[width=0.5\textwidth]{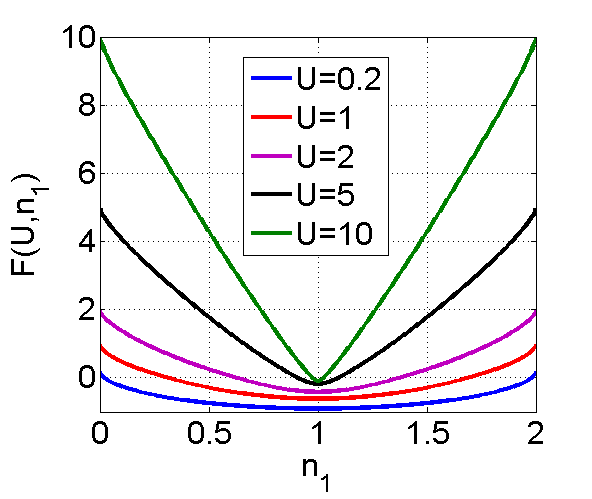}
 \caption{Universal part of the energy function(al) of a Hubbard dimer as a function of $n_1$ for several values of $U\!$.\, As $U$ increases, $F$ tends to $U|1{-}n_1|$. Reproduced from Ref. \cite{CFSB15}.}
 \label{Fn1}
\end{figure}

Although one can write analytic formulas for the ground-state energy for the dimer, there is no explicit analytic formula for $F\!$.\, It is trivial to calculate $F$ numerically and $F$ is shown in the Fig.~\ref{Fn1}. In the special case of $U=0$, it is easy,
\begin{equation}
    F_{U=0}(n_1) = T\s(n_1) = -\sqrt{n_1 (2 - n_1)} .
\label{FU0}
\end{equation}
Here we have attached the subscript S to remind us that $U=0$, so this is the kinetic energy function for a single Slater determinant, and is indistinguishable from the blue line of Fig.~\ref{Fn1}.

\ssec{Hohenberg-Kohn III}
\index{Hohenberg-Kohn!Theorem III}
HKIII states that there is a variational principle for the ground-state energy directly in terms of the density alone:
\begin{equation}
    E(\dd v) = \min_{n_1} \bigg\{ F_{\sss U}(n_1) + \dd v \dd n /2 \bigg\}.
\label{HK3}
\end{equation}
This bypasses all the difficulties of approximating the wavefunction (but of course buries them in the definition of $F_{\sss U}$). Usually, the minimum can be found from the Euler equation
\begin{equation}
    \frac{dF_{\sss U}(n_1)}{dn_1} - \frac{\dd v}{2} = 0,
\label{Euler}
\end{equation}
and the unique $n_1(\dd v)$ is the one that satisfies this equation.

This allows us to find a solution to the many-body problem, without ever calculating the wavefunction. Given an expression for $F_{\sss U}(n_1)$, either exact or approximate, for any value of $\dd v$, one can solve Eq.~\eqref{Euler} above to find the corresponding $\dd v$ (exact or approximate) and insert into Eq.~\eqref{HK3} to find the energy. Any approximation to $F(n_1)$ provides approximate solutions to all many body problems (every value of $\dd v$).

\begin{tcolorbox}
\textbf{Takeaway:} The HK theorems prove the existence of an exact variational principle for the ground-state energy based on the density, not the wavefunction, but give no information on how to approximate it. This is an (almost) useless statement in practice.  But to any unbeliever in DFT, one can always tell them (to go look at) $F_{\sss U}$.
\end{tcolorbox}

\sec{Kohn-Sham DFT}
\index{Kohn-Sham!density functional theory}\index{density functional theory!Kohn-Sham}
The original DFT, called Thomas-Fermi theory \cite{T27,F28}, tried to approximate $F_{\sss U}(n_1)$ directly, but such direct approximations have never been accurate enough for most electronic structure calculations. A tremendous step forward occurred when Kohn and Sham considered a fictitious system of non-interacting fermions with the same ground-state density as the true many-body one \cite{B12}. In our case, this is just the TB problem, for which we already have explicit solutions.

They wrote the $F$ function in terms of quantities that could easily be calculated in such a system:
\begin{equation}
    F_{\sss U}(n_1) = T\s(n_1) + U\H(n_1) + E\xc(n_1) .
\label{F_KS}
\end{equation}
Here, $T\s$ is just the TB hopping energy of Eq.~\eqref{dn_N2_U0}, and the Hartree energy is just the mean-field electron-electron repulsion
\begin{equation}
    U\H = \frac{U}{2}\big(n_1^2 + n_2^2\big),
\label{U_H}
\end{equation}
which is an explicit function of the occupations. Then $E\xc$, the exchange-correlation (XC) energy (about which, much more, later) is simply everything else, i.e., $E\xc$ is defined by Eq.~\eqref{F_KS}. It is then trivial to show, from the Euler equation, that the TB potentials that will reproduce the exact occupations are
\begin{equation}
    v_{\sss{S},i} = v_i + U n_i + \frac{\partial E\xc}{\partial n_i}\,.
\label{v_KS}
\end{equation}
The first correction to $v_i$ is the Hartree potential, while the second is the XC potential. These KS TB equations must be solved self-consistently, as the potentials depend on the occupations. Once converged, the final densities can be used to extract the total energy of the MB system, via
\begin{equation}
    E = T\s + U\H + E\xc + V = \epsilon - U\H + E\xc - \dd v\xc \dd n/2\,,
\end{equation}
where $\epsilon$ is the eigenvalue in the TB KS calculation. Again, just like in the HK case, once $E\xc(n_1)$ is given (either approximate or exact), the KS equations can be solved for any electronic system and a ground-state energy and occupation extracted.

The wondrous improvement due to the KS scheme is that only a small fraction of the total energy (the XC part) need be approximated. Many of the most important quantum effects, such as screening, shell structure, binding energies, etc.\ are mostly accounted for by the quantum effects of the one-body system. Finally, a very simple, intuitive approximation suggested by KS themselves (the local density approximation (LDA) \cite{D30,KS65}) produced far better results than they expected (but with binding energy errors too large for quantum chemistry taste).

\begin{figure}[t]
 \centering
 \vspace{-1ex}
 \includegraphics[width=0.49\textwidth]{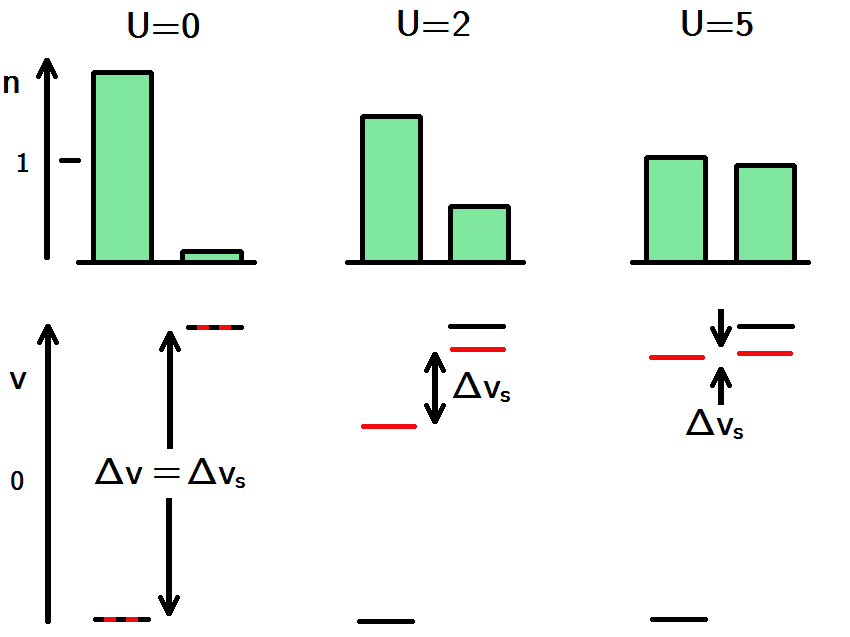}
 \vspace{-1ex}
 \caption{KS DFT view of an asymmetric half-filled Hubbard dimer as a function of $U$. The on-site potential difference $\dd v$ is shown in black and the KS on-site potential difference $\dd v\s$ is in red. Reproduced from Ref. \cite{CFSB15}.}
 \label{DFTdiag}
\end{figure}

Fig.~\ref{DFTdiag} gives us some sense of how this works, for $\dd v=1$. Then, if $U=0$, most occupation is on the left. For $U=2$, the repulsion makes the occupations more equal.  The KS potential is simply that TB potential that produces those (many-body) occupations.  So it must be a smaller potential difference than the real potential. One can see that the Hartree potential will typically overestimate repulsion, while XC corrects that to give the exact answer. Finally, when $U$ is ramped up to 5, the occupations become very close to equal, and the KS potential difference becomes very small.

Traditionally, $E\xc$ is separated into an exchange and a correlation contribution. The exchange contribution is then defined as
\begin{equation}
    E\x = \langle\mathrm{\Phi\s}|\hat V\ee|\mathrm{\Phi\s}\rangle - U\H,
\label{X}
\end{equation}
where $\Phi\s$ is the KS wavefunction, and $E\x$ is always negative. Then one can show correlation is just
\begin{equation}
    E\c = \langle\mathrm{\Psi}|\hat H|\mathrm{\Psi}\rangle - \langle\mathrm{\Phi\s}|\hat H|\mathrm{\Phi\s}\rangle
\label{C}
\end{equation}
and, by the variational principle, is also never positive. These definitions (almost) match those of quantum chemistry \cite{UG94}, except that in KS-DFT, all orbitals come from a single potential, while in HF orbitals are freely chosen to minimize the HF energy. But there are some surprises relative to the traditional many-body expansion. For example, because of the definitions, $E\x$ includes some `self-exchange', i.e., it is non-zero even for a single electron (where $E\x$ is $-U\H$ and $E\c = 0$). DFT approximations which do not satisfy these conditions for all one-electron densities are said to have self-interaction errors \cite{PZ81}. Moreover, `higher-order exchange effects' are all lumped into the correlation energy. In any event, for our 2-electron problem, in a spin singlet, $E\x = -U\H/2$, but no simple relation exists for larger $N\!$.

The traditional Hartree-Fock approximation comes from expanding the electron-electron interaction to first order, which means neglecting $E\c$, and then minimizing the energy. In full DFT terms, for our 2-electron system,
\begin{equation}
    F\HF = T\s + \half U\H,
\label{FHF}
\end{equation}
or in KS-DFT terms
\begin{equation}
    E\xc\HF = -U\H/2 \,.
\label{XCHF}
\end{equation}
Thus, solving the TB equation self-consistently with Eq.~\eqref{XCHF} produces the minimum for the total energy using $F\HF$ of Eq.~\eqref{FHF}.

\begin{figure}[t!]
 \centering
 \vspace{-2ex}
 \includegraphics[width=0.415\textwidth]{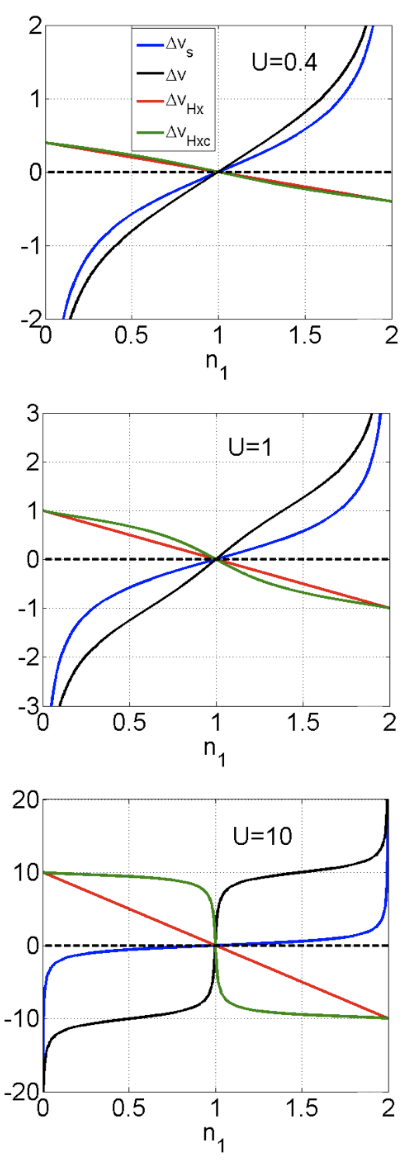}
 \vspace{-2ex}
 \caption{Plots of $\dd v\s$ (blue) and its components, the one-body potential $\dd v$ (black), the Hartree plus exchange potentials,  $U \dd n / 2$ (red), and the same with correlation added, $U \dd n /2 + \dd v\c$ (green) plotted against $n_1$ for various values of $U$. Reproduced from Ref. \cite{CFSB15}.}
 \label{dv}
\end{figure}
In Fig.~\ref{dv}, we show the contributions to the KS potential for a sequence of different $U$ values, as a function of the occupation. The effect of repulsion is to always oppose the potential difference, making the KS potential difference smaller. In the first, $U$ is small, and correlation is of order $U^2$ (see Reference \cite{CFSB15}). Thus the correlation contribution is negligible (red and green overlap) and HF is an excellent approximation. In the middle, $U=1$ is moderate, and now we begin to see the difference correlation makes in the potential. Moreover, its effect is to make $\dd v\Hxc$ deviate from a straight line. Finally, for strong correlation, the HXC potential (almost) exactly is equal and opposite to the one-body potential. Again, the HX contribution has much curvature, but now correlation wipes that out (almost) entirely. Clearly, the HF approximation will be terrible for the potential in this case, and yield entirely incorrect densities. In fact, a lower-energy solution appears if one allows spin symmetry breaking \cite{PSB95}.

It is now relatively routine to calculate accurate KS potentials from highly accurate densities found, e.g., via quantum chemical methods \cite{GDS12}. In an insanely demanding calculation, it is even possible to solve the KS equations using the exact XC functional \cite{WBSBW14}. Convergence becomes more
difficult as correlations grow stronger, but remains possible \cite{WSBW13}.

\begin{tcolorbox}
\textbf{Takeaway:} The KS scheme is exact meaning that, if we only knew the exact exchange-correlation functional, we could determine the ground-state energy exactly, of every electronic problem. There are many existing calculations of the exact XC potential. In practice, we must approximate XC, but because XC is a small fraction of the total energy, standard KS calculations are usefully accurate for ground-state energies and densities.
\end{tcolorbox}

\ssec{KS spectral function}
\index{Kohn-Sham!spectral function}
There is a {\em pernicious superstition} \cite{Tacitus} that the KS spectrum is related to the physical response properties of the real system. This false belief has arisen because, for weakly correlated systems, this is approximately true, apart from the fundamental gap of a semiconductor.
From a practical viewpoint, the KS bands are marvelously useful as a starting point for Green function calculations of real spectral functions.
Moreover, long ago, when the local density approximation ruled supreme, there was no way to know if differences between the KS and exact response properties was due to the crudeness of this approximation \cite{PPLB82,SS83}. These days, there are simple exact answers to such speculations, if we only have the patience to read them.

\ssec{The ionization potential theorem}
\index{ionization potential}
As a simple example of the mysterious workings of the exact functional, we state an important exact result
\begin{equation}
    I(N) = E(N{-}1) - E(N) = -\epsilon\HO (N) .
\label{IP}
\end{equation}
Here $E(N)$ is the ground-state of the $N$-electron system, and $\epsilon\HO$ is the energy of the highest-occupied KS orbital. (For those with some chemistry leaning, Koopmans' theorem is an approximate version of this for HF calculations \cite{PY89}). This illustrates some of the power of KS-DFT\@. You might think that, with the exact functional, all one can extract is the ground-state energy and density of our system. But the above result shows that the HO of the KS scheme also tells you the ionization energy. One can also extract all static response functions exactly by turning on weak external perturbations, and applying the exact functional to the perturbed systems. In practice, standard DFT approximations tend to violate this exact condition very badly \cite{PPLB82,SS83,PL83}. Nonetheless, they often still yield usefully accurate ground-state energies, thus performing their primary function. (On the other hand, returning to the discussion of HKI, knowing the exact XC does {\em not}, in general, give you access to, say, the first excited state energy.  It is a functional of $n_1$ alone, but we cannot deduce that functional from $E\xc(n_1)$.)

\begin{figure}[t]
 \centering
 \includegraphics[width=0.46\textwidth]{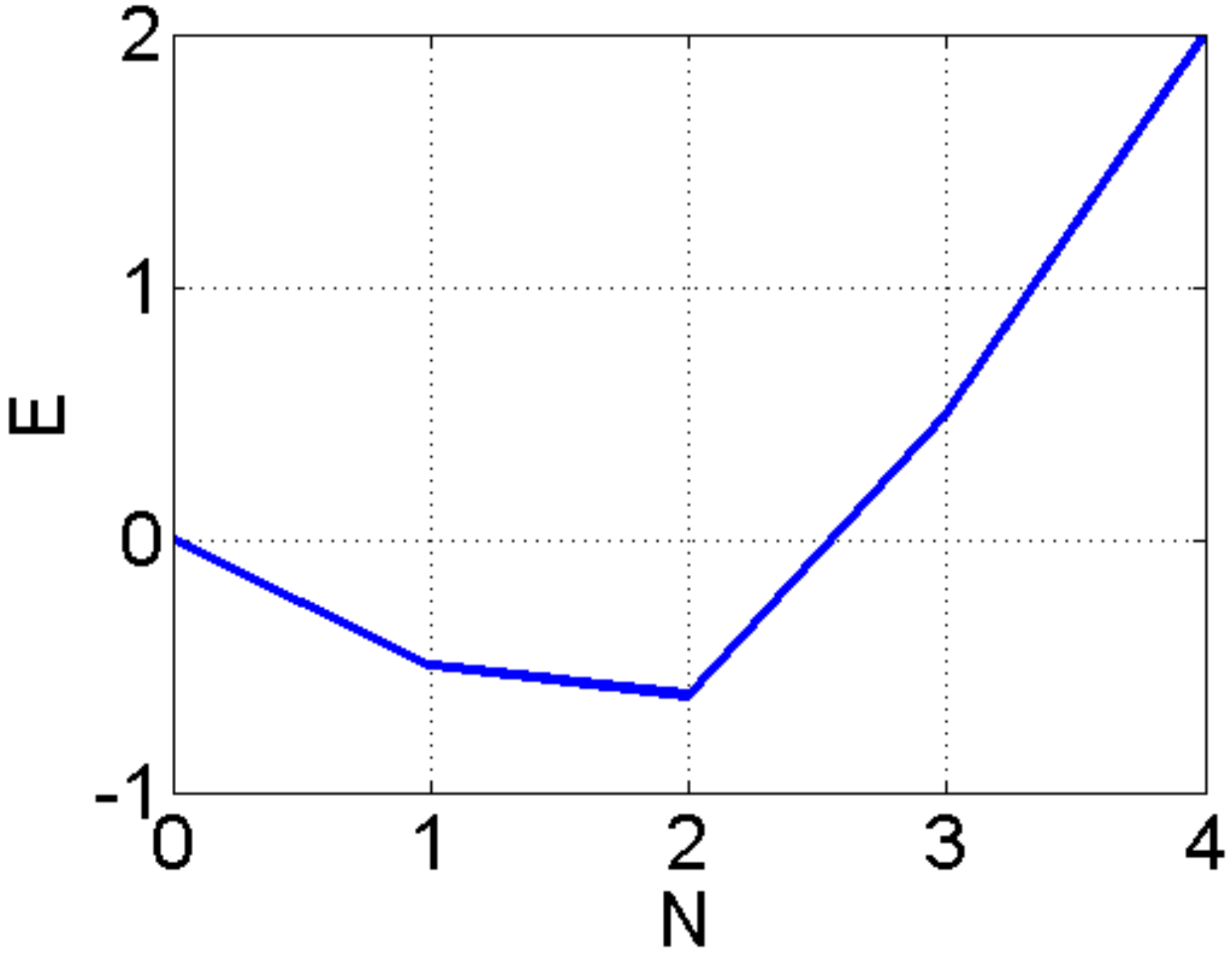}
 \caption{Plot of $E(N)$ for $U=1$ and $\dd v = 0$. Reproduced from Ref. \cite{CFSB15}.}
 \label{EN}
\end{figure}

Increasing $N$ by $1$ in Eq.~\eqref{IP} yields
\begin{equation}
    A(N) = E(N) - E(N{+}1) = -\epsilon\HO (N{+}1) \neq \epsilon\LU(N) ,
\label{EA}
\end{equation}
where $A$ is called the electron affinity of the system in chemistry. The difference between the KS HO of the $N{+}1$ electron system and the lowest unoccupied (LU) level of the $N$-electron system is called $\dd\xc$, where the $\dd$ indicates its origin from the infamous derivative discontinuity of DFT \cite{PPLB82}. This simply means, that at zero temperature, the energy of the system consists of straight line segments between integer values, as shown below in Fig.~\ref{EN}. The energy itself is continuous, but its derivative, the chemical potential, is not. For a neutral system, the chemical potential is $-I$ below the integer and $-A$ above. This discontinuous jump in $\mu$ shifts the KS HO eigenvalue by the same amount, producing the difference with the KS LU of the neutral. (Realistic electronic systems do not have an upward pointing portion of the curve in Fig.~\ref{EN}. This occurs for the dimer because electrons cannot escape to outside the system.)

\ssec{Mind the gap}
\label{sec_gap}
\index{band gap}

We are now ready to see the relevance of this to solids. Even for a finite system, we define the charge (or fundamental) gap as
\begin{equation}
    E_g = I - A \,.
\label{FG}
\end{equation}
As the size of the system grows toward a bulk material, this quantity tends to the fundamental charge (or transport) gap of the system (at least for ordered systems \cite{K64b}). But, because of Eqs.~\eqref{IP} and \eqref{EA} above, we find
\begin{equation}
    E_g = E_{s,g} + \dd\xc \,,
\label{KS_FG}
\end{equation}
where$E_{s,g}$ is the KS gap (i.e., the difference between the LU and HO level, or the gap between the KS valence and conduction bands in a solid). Thus, with the {\em exact} XC functional of ground-state DFT, we do not get the true gap by looking at its KS value for the neutral system.

\begin{figure}[t!]
 \centering
 \vspace{-3.5ex}
 \includegraphics[width=0.5\textwidth]{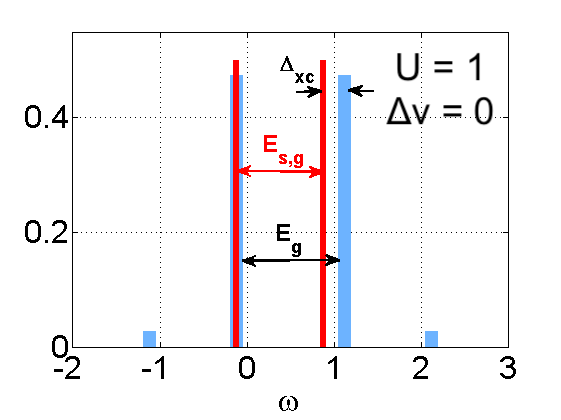}
 \vspace{-2ex}
 \caption{Spectral function of the symmetric dimer for $U=1$ and $\dd v = 0$. The physical MB peaks are plotted in blue, the KS in red. Here $I=0.1$, $A=-1.1$, and $\epsilon\LU=0.9$. Reproduced from Ref. \cite{CFSB15}.}
 \label{sf1}
\end{figure}

Fig.~\ref{sf1} shows the spectral function (projected onto the left-hand site) in a weakly correlated case \cite{CFMB18}, the symmetric dimer with $U=1$. We can see the sense in which the KS spectral function (red) resembles the blue exact one: the significant KS peaks are of about the same height and position as their blue counterparts, and the blue peaks without KS counterparts are relatively small. The KS gap is smaller than the true gap, but not by much.
Because both the KS and the exact spectral functions satisfy the same sum rule (even with an approximate XC), if the dominant peaks are reproduced (even with the wrong gap), only small peaks are missed in the KS spectrum.

\begin{figure}[t]
 \centering
 \vspace{-2.7ex}
 \includegraphics[width=0.5\textwidth]{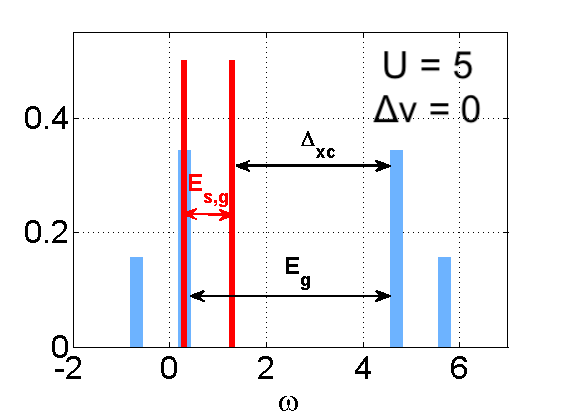}
 \vspace{-2.3ex}
 \caption{Same as Fig.~\ref{sf1}, but now $U=5$. Here $I=-0.3$, $A=-4.7$, and $\epsilon\LU=1.3$. Note that the KS gap remains unchanged by the alteration of $U$ because $\dd n=0$ in both cases. Reproduced from Ref. \cite{CFSB15}.}
 \label{sf2}
\end{figure}

On the other hand, Fig.~\ref{sf2} shows the same system with a larger $U$ value. Now the strong KS peaks are not in the right place and are noticeably too large. Moreover, the blue peaks with no KS analogs are a substantial contribution. 
\begin{figure}[t]
 \centering
 \vspace{-3ex}
 \includegraphics[width=0.5\textwidth]{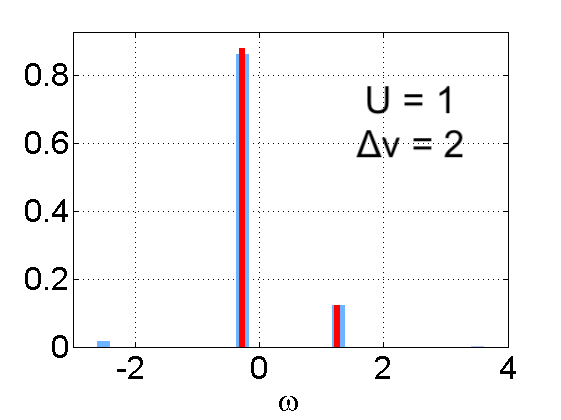}
 \vspace{-2ex}
 \caption{Same as Fig.~\ref{sf1}, but now  $U=1$, $\dd v = 2$. Here $I=0.27$, $A=-1.27$, and $\epsilon\LU =1.25$. Reproduced from Ref. \cite{CFSB15}.}
 \label{sf3}
\end{figure}
Finally, in the inhomogeneous case, the potential asymmetry overcomes the effects of the Hubbard $U\!$. In Fig.~\ref{sf3}, we see that for $\dd v = 2$ and $U = 1$, the KS spectral function is almost identical to the true one.

Lastly, we finish this section illustrating the relevance of this discussion to the thermodynamic limit. The canonical example of the Mott-Hubbard transition is a chain (or lattice) of H atoms. Each atom has one electron, so the bands of the KS potential are always half-filled, with no gap at the Fermi energy. Thus the gap is always zero and the KS band structure suggests it's a metal. This may be true at moderate separations of the atoms, but as the separation is increased, the electrons must localize on atoms, and it must become a Mott insulator.
\begin{figure}[t!]
 \centering
 \includegraphics[width=0.47\textwidth]{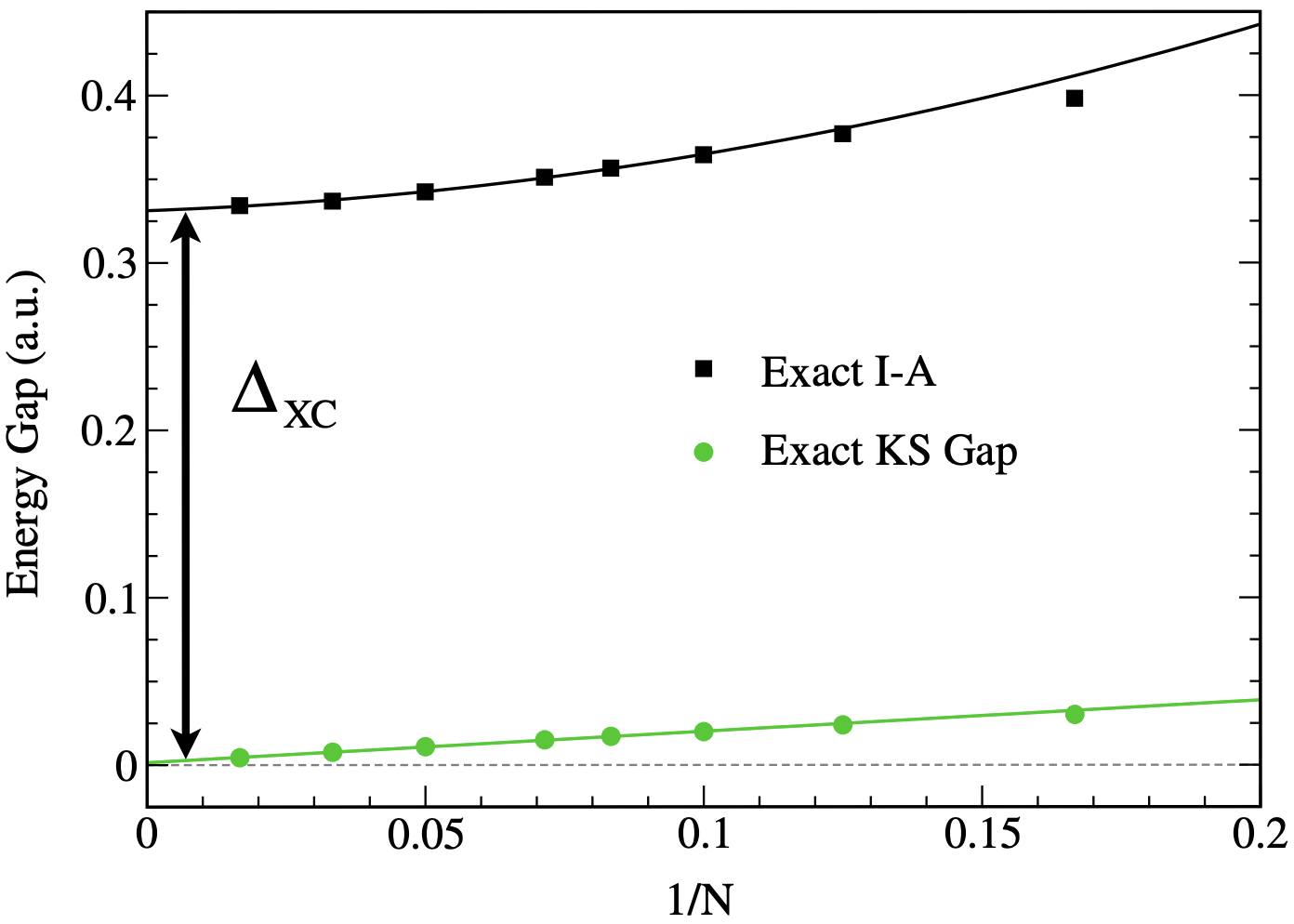}
 \vspace{-2ex}
 \caption{Exact gaps for chains of $N$ soft hydrogen atoms with atomic separation $b = 4$ (error bars are less than symbol sizes). The upper curve is a quadratic fit of exact gaps of the largest six systems and extrapolates to a finite value $E_g \approx 0.33$. The exact Kohn-Sham gaps, in contrast, extrapolate to zero showing that for $N\rightarrow\infty$ the true KS system is metallic (lower curve is a linear fit of exact KS gaps of the largest six systems). Reproduced from Reference \cite{SWWB12}.}
 \label{gaps}
\end{figure}

Fig.~\ref{gaps} shows the gap, calculated for chains of well-separated 1D H atoms of increasing length \cite{SWWB12}. By performing the calculation with finite systems, i.e., without periodic boundary conditions, we calculate the gap for each $N$ by adding and removing electrons, as in Eq.~\eqref{FG}, and then take the limit as $N\to\infty$. On the other hand, we extract the exact ground-state density from our DMRG calculation at each $N\!$,\, and find the corresponding exact KS potential for each $N\!$.\, We could then as easily extrapolate the KS gap, from the HO and LU, showing that indeed the KS gap vanishes in the thermodynamic limit --
exactly the same as if we had calculated the KS band structure, in which the Fermi energy would be right in the middle of the band.
This provides a dramatic illustration of the KS underestimate of the true gap, even when using the exact XC functional.

\begin{tcolorbox}
\textbf{Takeaway:} The KS Green function does not match the true Green function. If correlation is weak, it may be a good approximation to it, with its main deficiency being an underestimate of the gap. For stronger correlations, there can be huge differences, and there are always more features in the real Green function. In the thermodynamic limit, the exact KS gap can vanish for a simple Mott insulator.
\end{tcolorbox}

\ssec{Talking about ground-state DFT}

First, we review our crucial formal points.  
\begin{enumerate}
    \item In general, the KS scheme with the exact functional yields ground-state energy and density, and any other quantities that can be teased from them, such as static response properties and ionization potentials.
    \item There is no formal meaning for most KS eigenvalues in ground-state DFT, despite the fact that many practitioners treat them as if there were. Of course, they do provide tremendous physical and intuitive insight, especially for weakly correlated systems, where they are good approximations to the excitations (either quasi-particle or optical). But when correlations are strong, explicit methods are needed to correct them \cite{ZP19}.
    \item The strongest manifestation of point 2 above is that the exact KS gap is typically smaller than the true gap, and can vanish in cases where the true gap is finite (Mott insulator).
    \item Moreover, there is an exact formula relating the total energy to the sum of the KS eigenvalues, which contains finite corrections for double counting. There is no ambiguity about these corrections, they are derived from the formal theory, and yield the exact many-body energy.
    But when correlated methods are used for a \emph{subset} of the orbitals, ambiguities can arise that affect occupancies \cite{M12}.
    \item Although in principle, all properties are functionals of the ground-state density, knowledge of the exact ground-state energy functional (via $E\xc$) does not provide a way to calculate these other functionals.  As we see later, TDDFT is a way to do precisely this.
\end{enumerate}

Next, we discuss how these points show up in practical DFT calculations
of solids, where XC approximations must be made.

\begin{enumerate}
    \item The steady progress within quantum chemistry and materials in functional development is almost entirely focused on improvements in accuracy and reliability of the total energy for weakly correlated systems \cite{KBP96,H96}. This is by far the most important use of DFT in modern electronic structure.  Such improvements are often not particularly relevant to the response properties of greatest interest in strongly correlated materials. For example, the KS eigenvalues are often not improved significantly by functionals yielding better energies \cite{SH99,TH98}.
    Although the KS eigenvalues cannot be directly interpreted in general, they are uniquely defined (up to a constant). Thus the exact KS Hamiltonian is a well-defined starting point for many-body methods.
    \item The KS scheme is not a mean-field scheme in the traditional sense of the word, and it can be extremely difficult to relate its features to those of traditional many-body theory. The KS wavefunction is typically a single Slater determinant, but yields the exact many-body energy via its density.
    \item Standard approximations, such as LDA and generalized gradient approximations (GGA), by construction produce total energies that are smooth and continuous at integer $N$, unlike the exact $E(N)$. Thus their corresponding $\dd\xc$ is zero\cite{PPLB82}. According to Sec.~\ref{sec_gap}, the KS band gap in such approximations {\em is} their prediction for the fundamental gap. In fact, it has been found that their KS gaps are likely a good approximation to the exact KS gap \cite{GMR06}, but their lack of discontinuous behavior means they miss the correction to turn it into the true gap.
    \item On the other hand, the range-separated hybrid functional HSE06 is well-known to produce reasonable gaps for moderate gap semiconductors. This is because, instead of performing a true pure KS calculation, most codes (like VASP) perform a {\em generalized} KS calculation \cite{SGVML96} when a functional is orbital-dependent\cite{H08,Perdew17}. They treat the orbital-dependent part of the potential as if it were a many-body potential, just as is done in HF\@. (A similar but smaller effect occurs in meta-GGA's that depend on the kinetic energy density, such as SCAN \cite{SCAN}). And in fact clever tricks may be used to extract the true gap, even from a periodic code \cite{G15}.
\end{enumerate}
\begin{tcolorbox}
\textbf{Takeaway:} Even with the exact functional, the KS band gap does not equal the true transport gap of the system. Likely, semilocal functionals yield accurate KS gaps but, because they lack a discontinuous behavior at integer particle numbers, cannot yield accurate transport gaps. Modern hybrid functionals that depend explicitly on KS orbitals yield band gaps closer to fundamental gaps, but only when treated with generalized KS theory.
\end{tcolorbox}

\sec{Time-dependent DFT (TDDFT)}\label{SecTDDFT}
\index{time-dependent density functional theory}
Our last main section is about time-dependent density functional theory (TDDFT) \cite{BWG05,MMNG12,U12,Maitra16}. While this uses many of the forms and conventions of ground-state DFT, it is in fact based on a very different theorem from the HK theorems. When applied to the linear response of a system to a dynamic electric field, it yields the optical transitions (and oscillator strengths) of that system. It has become the standard method for extracting low-level excitations in molecules, where traditional quantum chemical calculations are even more demanding than those for the ground state.

The Runge-Gross theorem \cite{RG84} states that, for a given initial wavefunction, statistics, and interaction, the {\em time-dependent} density uniquely determines the one-body potential. In principle, this can be used for any many-electron time-dependent problem, including those in strong laser fields \cite{BWG05}. In practice, such calculations are limited by the accuracy of the approximations and whether the observable of interest can be extracted directly from the one-electron density. One constructs TD KS equations, defined to yield the exact time-dependent one-electron density. Because TDDFT applies to the time-dependent Schr\"odinger equation, the XC functional differs from that of ground-state DFT in general, and has a time-dependence.

Our interest will be only in the linear-response regime. In that case, one can derive a crucial result, which we give in operator form, called the Gross-Kohn equation \cite{GK85}
\begin{equation}
    \chi (\omega) = \chi\s (\omega) + \chi\s (\omega) * (f\H + f\xc (\omega)) * \chi (\omega) ,
\label{GK}
\end{equation}
where $\chi (\omega)$ is the dynamic density-density response function of the system, and $\chi\s$ is its KS counterpart. The kernel, $f\!$,\, is the functional derivative of the time-dependent potential. Thus, $f\H$ is the Hartree contribution, while $f\xc(\omega)$ is the XC correction.   

Eq.~\eqref{GK} is a Dyson-like equation for the polarization. If we set $f\xc=0$, it is the standard random-phase approximation, the Coulomb interaction simply dressing the bare interaction, and producing all the bubble diagrams. But things get a little weird when we assert that inclusion of $f\xc(\omega)$ produces the {\em exact} response of the system, for all frequencies. From a many-body viewpoint, this is suspicious, as these are a closed set of equations without coupling to 4-point functions. But the logic is sound and exactly analogous to the ground-state: there exists such a function that could be considered as defined by Eq.~\eqref{GK}.

The excitations of a system are given by poles of its response function. Simple analysis (exactly that of RPA) yields a matrix equation that corrects KS transition frequencies to the true transition frequencies, where the matrix elements involve $f\H + f\xc$. With standard ground-state approximations, folks have merrily calculated mostly low-lying valence transitions from the ground-state of many molecules \cite{AJ13}, finding accuracies a little lower than those of ground-state DFT \cite{JWPA09}, and computational costs that are comparable. This has been invaluable for larger molecules, where many excitations of the same symmetry may overlap, and so TDDFT yields a semiquantitative signature that can be easily matched with experiment \cite{BAHK98}.   

However, not all is well in paradise.  Almost immediately, it was noticed that the use of a ground-state approximation is simply the static limit of the corresponding kernel, and can be easily shown to produce only single excitations. While useful workarounds were created for some cases, it was also found that going to higher-order response does not solve the problem. And many of the most exciting transitions in biochemistry are double excitations.

\begin{tcolorbox}
\textbf{Takeaway:} Time-dependent DFT applies DFT methods to time-dependent problems. Within linear response, this yields exact expressions for the dynamic polarization, but at the cost of introducing a new functional, the frequency-dependent XC kernel. Ignoring its frequency dependence yields useful accuracy for low-lying molecular excitations with standard functionals. TDDFT is now standard for calculating optical response of molecules and materials.
\end{tcolorbox}

\begin{figure}[t]
 \centering
 \vspace{-1.5ex}
 \includegraphics[width=0.45\textwidth]{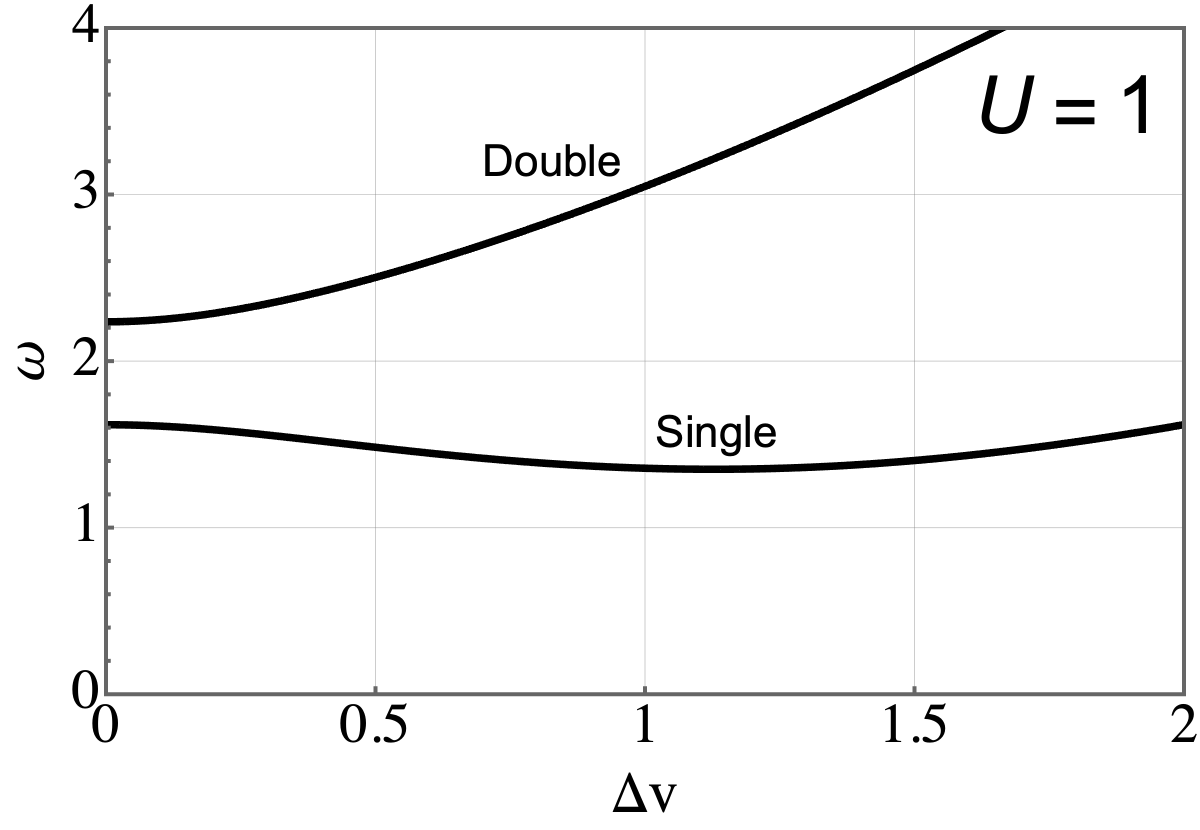}
 \vspace{-2ex}
 \caption{Transition frequencies of the first and second excitations as a function of $\Delta v$ for $U=1$.} 
 \label{TDDFT1}
\end{figure}

\ssec{Hubbard dimer}
\index{Hubbard dimer}
Happily we care only about Hubbard dimers, where everything is much simpler. First, we note our Hubbard dimer, in the singlet space, has just three states: the ground-state, the first excited state, which has a single excitation, and the second excited state, which is a double excitation out of the ground-state. Since there are no spatial degrees of freedom, our $\chi (\omega) $ is the Fourier transform of $\dd \n(t)/\dd v(0)$, which is just a scalar, with $\omega$-dependence
\begin{equation}
    \chi_n(\omega) = \frac{a_1}{\omega^2-\omega_1^2} + \frac{a_2}{\omega^2-\omega_2^2},
\label{HDRes}
\end{equation}
where $\omega_i$ denotes the transition frequency and $a_i$ is related to its oscillator strength \cite{CFMB18}. Thus $\chi$ has poles at each of the transition frequencies. Fig.~\ref{TDDFT1} shows the value of each of these transitions as a function of $\dd v$ for $U=1$. The double excitation is a little above the single for the symmetric case, but grows linearly with $\dd v$. The single remains about the same, and even dips, until $\dd v=U\!$,\, and then begins to grow itself. Here we can use our model system to examine one of the key mysteries of practical TDDFT: Where did all the higher excitations go?

\begin{figure}[t]
 \centering
 \vspace{-2ex}
 \includegraphics[width=0.45\textwidth]{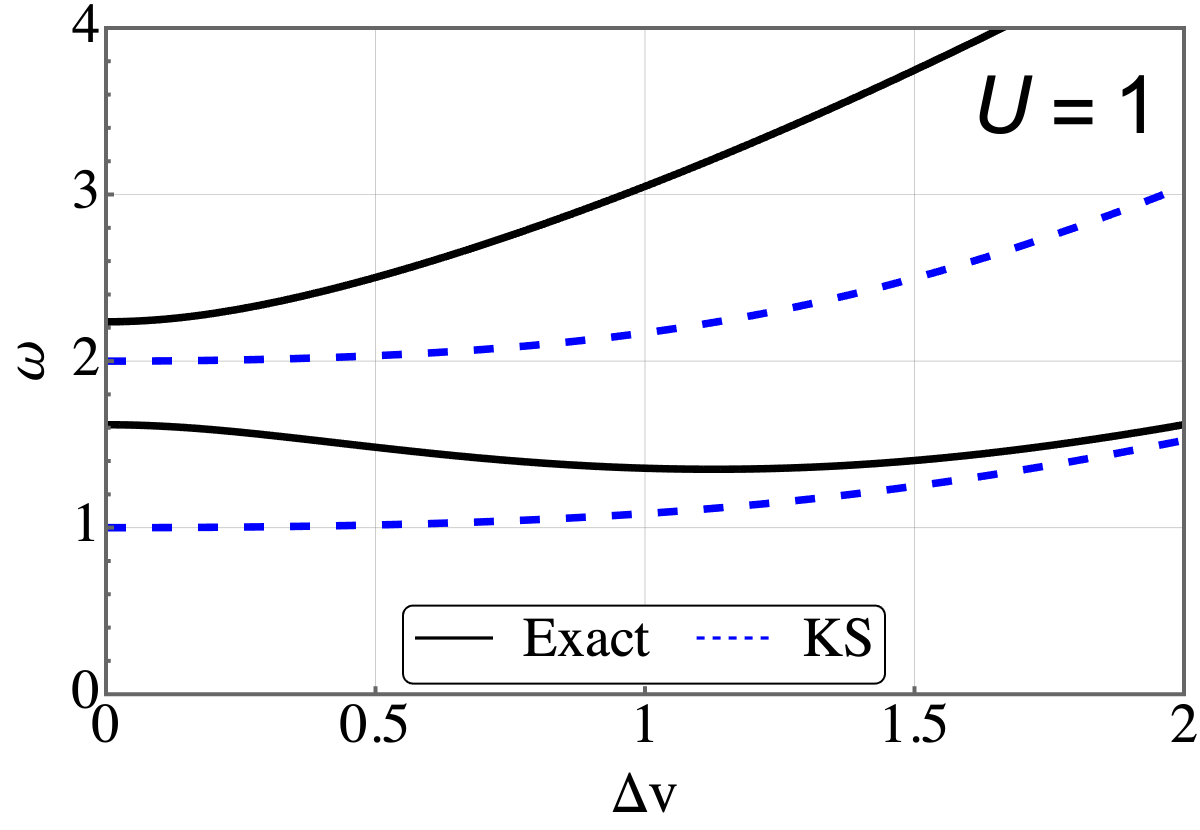}
 \vspace{-2.5ex}
 \caption{Same as Fig.~\ref{TDDFT1}, but with KS transitions (depicted in blue). For $\dd v>U\!$, the KS transition is a very good approximation to the true transition.}
 \label{TDDFT2}
\end{figure}

First we do an exact ground-state KS calculation, as in the previous sections. Thus the exact KS system is a tight-binding problem with effective potential, $\dd v\s$, defined to yield the exact ground state $\dd n$. This yields two eigenvalues, the lower symmetric combination and the higher asymmetric combination. The KS ground-state has the lower one doubly occupied. There do exist KS analogs of the many-body states. The single excitation has one electron excited to the higher level, the double has both. Fig.~\ref{TDDFT2} adds the KS transitions to Fig.~\ref{TDDFT1}, showing that they loosely follow the accurate transitions, but are significantly different. 

In the KS response function, $\chi\s$, the matrix elements of the density operator between ground and double excitation are zero, since both KS orbitals are different, so the Slater determinants are not coupled by a single density operator. Hence, such states have no numerator, eliminating any poles that might have arisen in the denominator, i.e.,
\begin{equation}
    \chi\s(\omega) = \frac{a_s}{\omega^2-\omega_s^2} \,.
\end{equation}
Thus the second KS transition, the double, does not appear at all in the response function! It's position is correctly marked in Fig.~\ref{TDDFT2}, but cannot be seen in $\chi\s$.

\begin{figure}[t]
 \centering
 \vspace{-1ex}
 \includegraphics[width=0.45\textwidth]{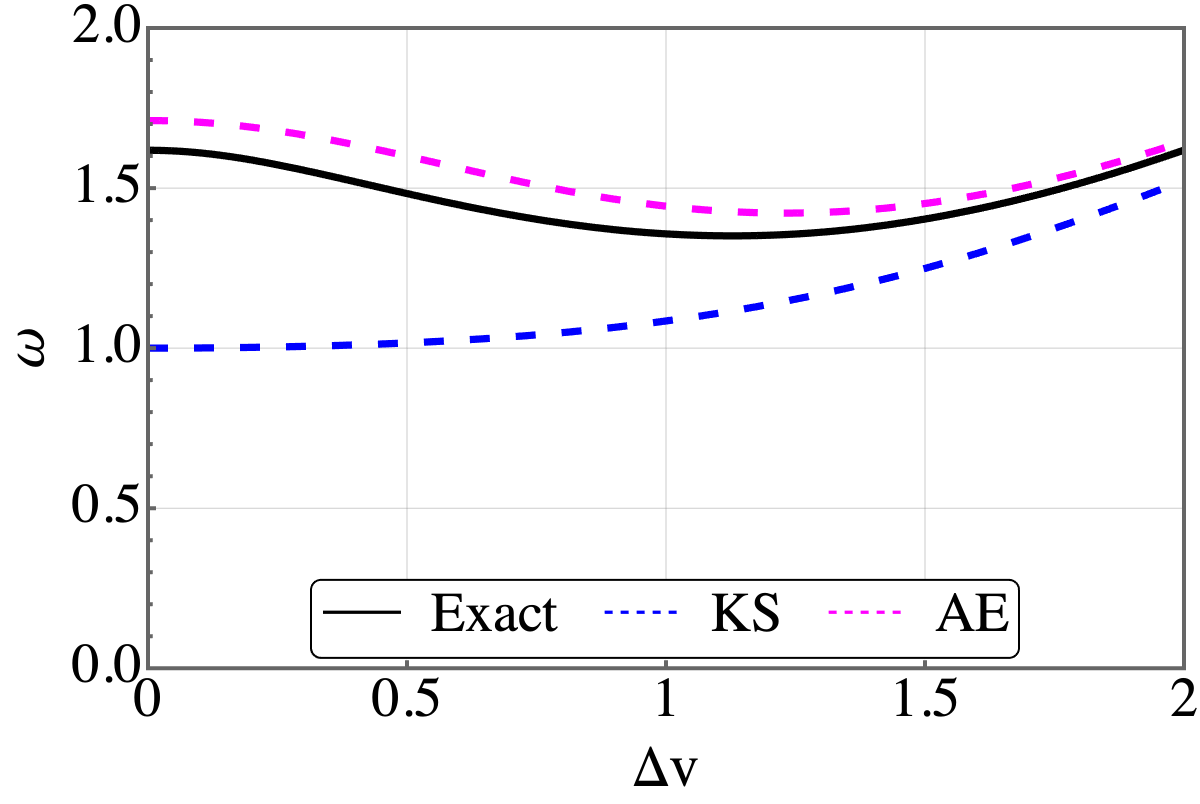}
 \vspace{-2ex}
 \caption{Same as Fig.~\ref{TDDFT1}, but with the adiabatically exact approximation (AE, pink dashes).} 
 \label{TDDFT3}
\end{figure}

By requiring the poles occur at the right places, one finds (in general) a matrix equation in the space of single excitations for the true transitions, whose elements are determined by the kernel. Here, this is one dimensional, yielding
\begin{equation}
    \omega^2 = \omega\s^2 + 2 \omega\s\, f\Hxc(\omega)\, \frac{2}{1+\dd v\s^2} \,.
\label{freq}
\end{equation}
The adiabatically exact approximation (AE) is to use the exact ground-state functional here to calculate $f\Hxc$. This corrects the single KS transition and is shown in Fig.~\ref{TDDFT3}. This works extremely well to capture almost all the difference with the KS transition, yielding very accurate excitations. This becomes even better for $\dd v$ greater than $U$, where the corrections virtually vanish (just as in Fig.~\ref{sf3} for the spectral function).

\begin{figure}[t!]
 \centering
 \vspace{-3ex}
 \includegraphics[width=0.45\textwidth]{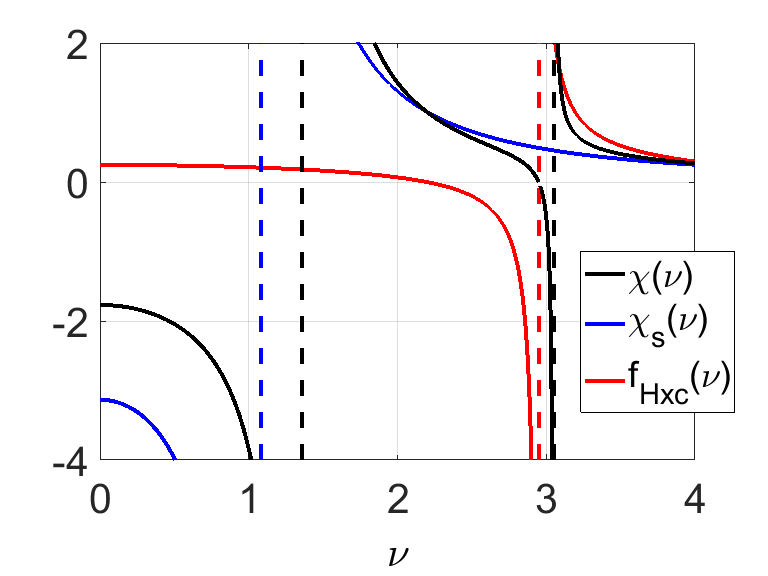}
 \vspace{-2ex}
 \caption{Frequency dependence of exact (black) and Kohn-Sham susceptibilities (blue) and exchange-correlation kernel (red) for $U = \dd v = 1$. Poles marked by dashed vertical lines, as a function of frequency $\nu$. The red line shows the exchange-correlation kernel. Reproduced from Ref. \cite{CFMB18}.}
 \label{kernel}
\end{figure}

But Eq.~\eqref{freq} just has one solution if the $\omega$-dependence in the kernel is neglected. On the other hand, if there is strong frequency dependence in the kernel, new transitions, not in the KS system, may appear. In fact, we know that is precisely what happens, as the physical system {\em does} have a double excitation. To understand how standard TDDFT fails, we note that we can calculate the exact kernel by finding $\chi(\omega)$ from many-body calculations, $\chi\s(\omega)$ by the techniques of the earlier section, inverting and subtracting
\begin{equation}
    f\Hxc(\omega) = \chi\s^{-1}(\omega) - \chi^{-1}(\omega) \,.
\label{fHxc}
\end{equation}
Fig.~\ref{kernel} shows the singular frequency-dependence of the kernel from Eq.~\eqref{fHxc}, which allows Eq.~\eqref{freq} to have an additional solution.

However, while all this provides insight into how the exact functional performs its magic, it does not tell us directly how to create a general purpose model, which would build this frequency-dependence into an explicit density functional sufficiently accurately to capture double excitations \cite{Maitra16}.

\begin{tcolorbox}
\textbf{Takeaway:} The Hubbard dimer demonstrates the accuracy of the adiabatic approximation in TDDFT
for single excitations, and also the missing frequency dependence needed to generate the double
excitations missing in adiabatic TDDFT\@.
\end{tcolorbox}

\ssec{Talking about TDDFT}

We saw in the earlier sections how the KS eigenvalues did not have a formal meaning in pure ground-state KS-DFT\@. We have seen here that, with the advent of TDDFT, they form the starting point of a scheme which produces the {\em optical} excitations. These are not the quasi-particle excitations associated with the Green function, which involve a change in particle number.

While the primary function of approximate ground-state DFT is to find energies, it usually also produces reasonably accurate densities, but rather erroneous XC potentials. In fact, this feat is achieved by having all the occupied orbitals shifted (higher) than their exact KS counterparts.  A constant shift has no effect on the density. But if the unoccupied levels (at least, the low-lying valence excitations) suffer the same shift, then KS transition frequencies are unaffected, and the adiabatic approximation (usually applied to the same XC approximation as the ground-state calculation) is reasonably accurate for many weakly correlated molecules.

Linear-response TDDFT has been less used in solids, because in the case of insulators, it became clear early on \cite{ORR02} that there is a long-range contribution to the XC kernel (as long-ranged as the Hartree contribution is) that is missed when using a semilocal ground-state approximation adiabatically. There are now many ways around this difficulty \cite{SDSG11}, some based on modelling the kernel using many-body techniques.

There have been many other approaches suggested for extracting optical excitations from DFT\@. An old simple one is called $\dd$-SCF \cite{ZRB77}, which involves simply using excited-state occupation numbers in a KS calculation, and finding the energy the usual way. Another, which has seen considerable recent interest \cite{GP19,YPBU17}, is to use ensemble DFT \cite{GOK88}.  

\begin{tcolorbox}
\textbf{Takeaway:} TDDFT can be considered an algorithm for finding the functional (of the ground-state density) for optical excitations.
\end{tcolorbox}

\sec{Summary}

This short review is aimed at broadening understanding of the basic differences between a density functional viewpoint and that of traditional many-body theory. The emphasis here has been on the exact theory, which we have illustrated on the 2-site Hubbard model. We have shown it is confusing to consider KS theory as any kind of traditional mean-field theory, and how the addition of TDDFT allows one to consider the KS eigenvalues as zero-order approximations to the optical excitations, not the quasiparticle excitations.  

However, the only reason that anyone cares about the exact theory of DFT is because, in practice, it is extremely useful with relatively unsophisticated approximations. These begin with the famous local density approximation, in which the XC energy per electron at each point in a system is approximated by that of a uniform gas matching the density at this point.  This was introduced already in the KS paper (where the statement of exactness appears as a mere footnote), thereby totally muddying the waters between exact and approximate statements. 
Walter Kohn told KB that he simply noticed the exact nature of the KS scheme after submitting the paper.
From about 1990 onwards, many users began using more sophisticated functionals, whose primary effect was to improve total energies and energy differences.

This article has said little or nothing about how to understand such approximations. This is because local (and semilocal) approximations capture a universal limit of all electronic systems, by yielding relatively exact XC energies in this limit \cite{LS73,EB09,CCKB18,OB21}. Traditional many-body theory generally considers a power series expansion in the electron-electron interaction. The alternative limit simultaneously increases the number of particles, in a way that the total electron-electron repulsion remains a finite fraction of the total energy even as interactions become weaker. The simplest example of this is that the LDA for exchange, whose formula can be derived by hand, has a percentage error that vanishes for atoms as $Z=N\to\infty$ \cite{EB09}.

This limit is hard-wired into the last term of the real-space Hamiltonian of Eq.~\eqref{ham1}, which is the integral of the density times the one-body potential. This is why the density is the basic variable in DFT\@. Even if formal theorems can be proven using other variables, this is why density functional theory has been so successful. It is also the case that the one-body potentials to which we apply DFT are diagonal in coordinate space, which is related to why the LDA is a universal limit.

Thus, key aspects of DFT approximations that are crucial to its success are missing from lattice models like the Hubbard model. There is no corresponding universal limit in which LDA becomes exact, even if one uses an approximation based on the uniform case \cite{LOC02,FVC11}. Again, this is why we created our 1D real-space mimic of 3D reality, instead of just solving lattice models.

\begin{tcolorbox}
\textbf{Takeaway:} This chapter has illustrated a variety of key conceptual points about DFT on a simple model system. Anyone who can answer the exercises will have absorbed 90\% of the material, and should be well-qualified to understand exactly what a DFT calculation does, and does not, tell you. In the twenty-first century, with so many DFT calculations being performed in so many different fields, the phrase ``Oh, that's just mean-field theory'' should no longer have any place in scientific discussions about DFT results. 
\end{tcolorbox}

\sec{Acknowledgments}
K.B.~acknowledges support from the Department of Energy, Award No. DOE DE-SC0008696. J.K.~acknowledges support from the Department of Energy, Award No. DE-FG02-08ER46496. We thank Eva Pavarini for suggesting this chapter, and making us write it.

\newpage
\appendix
\sec{Exercises}

If you have followed the logic throughout this tutorial, you will enjoy sorting out these little questions. If you want solutions, please email either of the authors, with a brief note about your current status and  interests.

\begin{enumerate}
\item
State which aspect of Fig.~\ref{dvdn} illustrates the HKI theorem.

\item 
What geometrical construction gives you the corresponding ground-state potential for a given $n_1$ in Fig.~\ref{Fn1}?

\item
Study the extreme edges ($n_1=0$ and $2$) of Fig.~\ref{Fn1}. What interesting qualitative feature is barely visible, and why must it be there?

\item
What feature must always be present in Fig.~\ref{Fn1} near $n_1=1$?  Explain.

\item
How can you be sure that, no matter how large $U$ becomes, $F_{\sss U}(n_1)$ is never quite $U|1{-}n_1|$?

\item
Assuming the blue line is essentially that of $U=0$, use geometry on Fig.~\ref{dndv} to find $\dd v\s$ for $U=5$.

\item
What is the relation, if any, between each of the blue plots in the three panels of Fig.~\ref{dv}? Explain.

\item
What is the relation, if any, between each of the red plots in the three panels of Fig.~\ref{dv}? Explain.

\item 
Why is the green line almost the mirror image of the black line in the $U=10$ panel of Fig.~\ref{dv}? Could it be the exact mirror image? Explain.

\item
From Fig.~\ref{EN}, using $E(N)$ about $N{\;=\;}2$, determine the locations of the largest peaks of Fig.~\ref{sf1} and compute the gap between them.

\item
Sketch how Fig.~\ref{EN} must look if $U=10$ and $\dd v=0$.  

\item
What is the relation between the two blue lines in Fig.~\ref{TDDFT2}? Explain.

\item
Give a rule relating the numbers of vertical lines of different color in Fig.~\ref{kernel}.\\ Explain its significance.

\item
Recall the definition of the kernel from section \ref{SecTDDFT}. Using this, derive $f\H$ and $f\x$, and draw them on Fig.~\ref{kernel}. Explain where double excitations must come from for 2 electrons.

\item 
Using formulas and figures from both sections, deduce the results of Fig.~\ref{TDDFT3} in the absence of correlation (Hint: You will need to solve the Hartree-Fock self-consistent equations), and comment on the relative errors. This is a little more work than the other exercises.

\end{enumerate}

\bibliography{Master}

\label{page:end}
\end{document}

%% file: Burke_template.tex
\usepackage{amsmath}
\usepackage{physics}
\usepackage{amssymb}
\usepackage{float}
\usepackage[dvipsnames]{xcolor}

\definecolor{TITLECOL}{rgb}{0.05,0.25,0.85}
\definecolor{CONTENTSCOL}{rgb}{0.1,0.2,0.7}
\definecolor{URLCOL}{rgb}{0,0.52,0.83}
\definecolor{LINKCOL}{rgb}{0.05,0.5,0}
\definecolor{CITECOL}{rgb}{0.25,0,0.48}
\definecolor{SECOL}{rgb}{0.07,0.31,0.80}
\definecolor{SSECOL}{rgb}{0.26,0.19,0.75}

\newcommand{\coloredtitle}[1]{\title{\textcolor{TITLECOL}{#1}}}
\newcommand{\coloredauthor}[1]{\author{\textcolor{CITECOL}{#1}}} 
\renewcommand{\sec}[1]{\section{\textcolor{SECOL}{#1}}}
\newcommand{\ssec}[1]{\subsection{\textcolor{SSECOL}{#1}}}

\usepackage{graphicx}

\def\PublicationsLink{http://dft.uci.edu/publications.php}
\def\preprintlink{ \href{\preprintlinklocation}{\TitleOfPaper} }
\def\preprinttext{~}

\usepackage{fancyhdr}
\makeatletter
\fancypagestyle{titlepage}
{	\lhead{\textsc{\preprinttext\ ~ \href{\PublicationsLink}{~}}}
	\chead{}
	\rhead{\preprintlink}
	\lfoot{}}
\makeatother
\def\preprintlink{ 
	\href{\preprintlinklocation}
        {
~}
	}

\definecolor{Green}{rgb}{0.016,0.627,0}
\definecolor{Plum}{rgb}{0.17,0,0.45}
\definecolor{LBlue}{rgb}{0,0.34,0.45}
\definecolor{Sepia}{rgb}{0.37,0.17,0.02}
\definecolor{BurntOrange}{rgb}{0.78,0.39,0}

\usepackage{hyperref}
\hypersetup{
    breaklinks=true,
    bookmarksopen=true,
    bookmarksnumbered=true,
    colorlinks=true,
    linkcolor=LINKCOL,
    linktocpage=true,
    citecolor=CITECOL,
    filecolor=magenta,      
    urlcolor=URLCOL,
}

%% file: KieronsMacros.tex
\def\bea{\begin{eqnarray}}
\def\eea{\end{eqnarray}}
\def\ben{\begin{equation}}
\def\een{\end{equation}}
\def\benu{\begin{enumerate}}
\def\enu{\end{enumerate}}

\def\bei{\begin{itemize}}
\def\eei{\end{itemize}}
\def\beit{\begin{itemize}}
\def\eit{\end{itemize}}
\def\benu{\begin{enumerate}}
\def\enu{\end{enumerate}}

\def\n{n}

\def\sss{\scriptscriptstyle\rm}

\def\hatH{{\hat H}}
\def\1var{(\bx_1...\bx\N)}

\def\half{\frac{1}{2}}

\def\bx{{x}}

\def\x{_{\sss X}}
\def\c{_{\sss C}}
\def\s{_{\sss S}}
\def\xc{_{\sss XC}}

\def\Hxc{_{\sss HXC}}

\def\N{_{\sss N}}
\def\H{_{\sss H}}

\def\HF{^{\rm HF}}

\def\HO{^{\rm HO}}
\def\LU{^{\rm LU}}

\def\ee{_{\rm ee}}

\def\sph_int{ {\int d^3 r}}

\def\dd{\mathrm{\Delta}}